\documentclass[aps,twocolumn,showpacs,floatfix]{revtex4}
\usepackage[T1]{fontenc}
\usepackage{graphicx}
\usepackage{amstext}
 \usepackage{amsxtra}

\newcommand\forget[1]{}

\begin{document}

%\preprint{}

%\title{Ultranarrow CPO effect in a $\Lambda$-system}
\title{Ultranarrow CPO resonance in a $\Lambda$-type atomic system}



\author{T. Laupr\^etre$^1$}
\author{S. Kumar$^{2}$}
\author{P. Berger$^{1,3}$}
\author{R. Faoro$^1$}
\author{R. Ghosh$^{2}$}
\author{F. Bretenaker$^1$}
\author{F. Goldfarb$^1$}
\email{Fabienne.Goldfarb@u-psud.fr}
\affiliation{$^1$Laboratoire Aimé Cotton, CNRS-Université Paris Sud 11, 91405 Orsay Cedex, France}
\affiliation{$^2$School of Physical Sciences, Jawaharlal Nehru University, New Delhi 110067, India}
\affiliation{$^3$Thales Research and Technology, Campus Polytechnique, 91767 Palaiseau Cedex, France}

\date{\today}
\begin{abstract}
It is well known that ultranarrow electromagnetically induced transparency (EIT) resonances can be observed in atomic gases at room temperature. We report here the experimental observation of another type of ultranarrow resonances, as narrow as the EIT ones, in a $\Lambda$-system selected by light polarization in metastable $^{4}$He at room temperature. It is shown to be due to coherent population oscillations in an open two-level system (TLS). For perpendicular linearly polarized coupling and probe beams, this system can be considered as two coupled open TLSs, in which the ground state populations exhibit anti-phase oscillations. We also predict theoretically that in case of two parallel polarizations, the system would behave like a closed TLS, and the narrow resonance associated with these oscillations would disappear.
\end{abstract}

%\pacs{}
%\keywords{}

\maketitle

Coherent population oscillations (CPO) and electromagnetically induced transparency (EIT) are two phenomena that can give rise to resonances much narrower than the relaxation rate of optical coherences. The former happens in two-level atomic systems (TLS), when the beatnote between a coupling beam and a coherent probe beam leads to a temporal modulation of the population difference. The width of the induced transparency window is then limited by the population relaxation rate \cite{Schwartz, Boyd}. The latter is a two-photon phenomenon that occurs for example in three-level $\Lambda$ systems, when two optical transitions couple two lower levels to a common upper one. When coherent laser beams excite both transitions, a narrow transparency window appears at Raman resonance, the width of which is limited by the Raman coherence lifetime \cite{Boller}. As the Raman coherence lifetime can be much longer than the upper level population lifetime, EIT usually leads to the narrowest resonances. In the last two decades, such phenomena have raised a lot of interest as they allow one to reach very slow group velocities for light \cite{Hau,Kash,Bigelow}, EIT being even used for light storage experiments \cite{Liu, Phillips, Camacho,Vudyasetu}.

In this Letter, we report the experimental observation of ultranarrow resonances in the absorption spectrum of a hot atomic vapor. These resonances, that cannot be attributed to EIT, are shown theoretically and experimentally to be due to CPO in the two coupled TLSs provided by the $\Lambda$ system. The two ground state populations exhibit anti-phase oscillations, while the total population is conserved. We compare all the features of these CPO resonances with those of the EIT ones.

\begin{figure}[h]
  \includegraphics[width=1.0\linewidth]{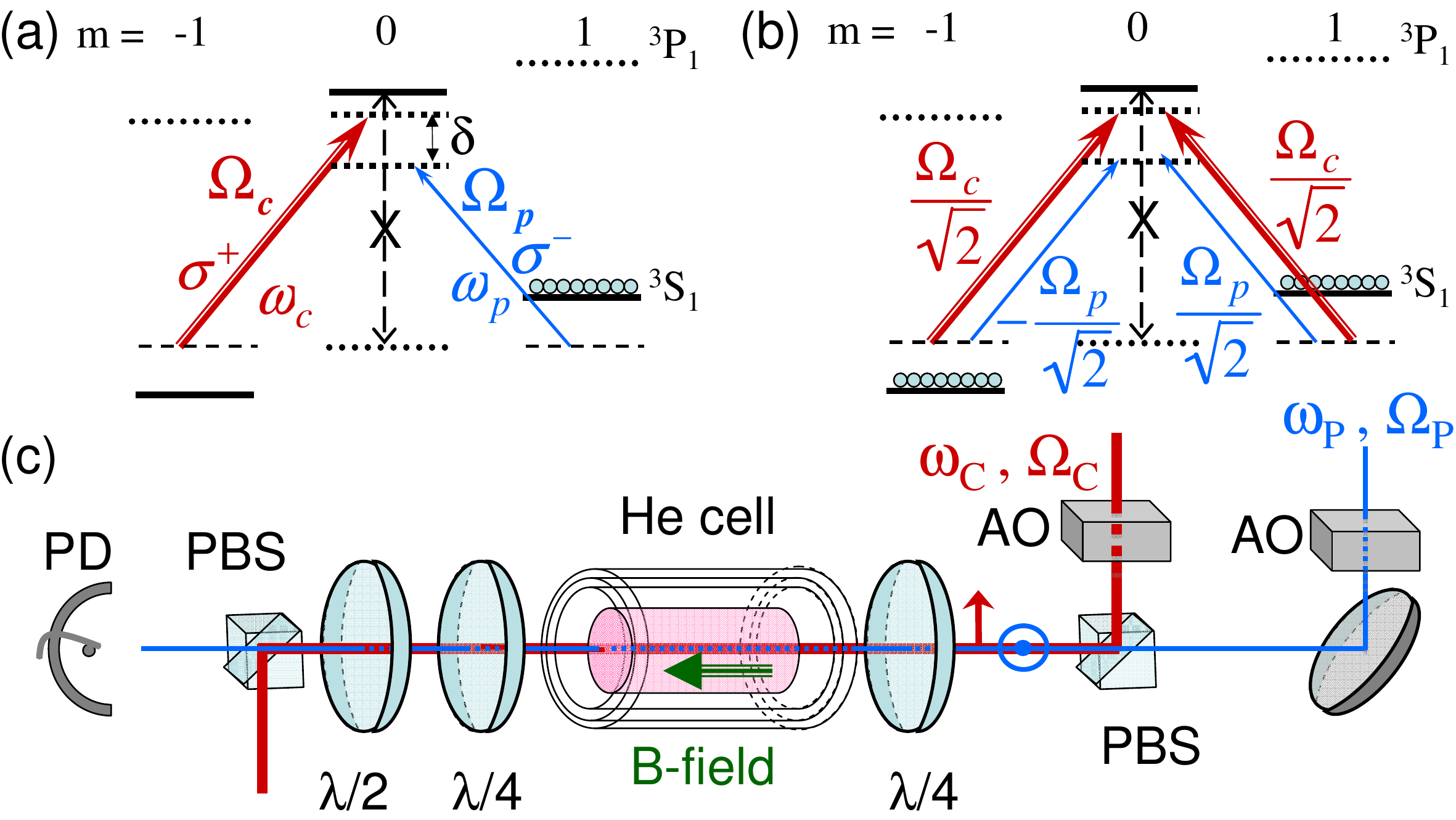}
  \caption{(color online) (a) and (b) Relevant level schemes in the case of excitation by orthogonal circular ($\sigma\perp\sigma$) and linear ($\mathrm{lin}\perp\mathrm{lin}$) polarizations. $\Omega_{\mathrm{P}}$ ($\Omega_{\mathrm{C}}$) and  $\omega_{\mathrm{P}}$ ($\omega_{\mathrm{C}}$): Rabi and optical frequencies of the probe (coupling) beam. $\delta=\omega_{\mathrm{p}}-\omega_{\mathrm{c}}$. (c) Experimental setup. AO: acousto-optic modulator. PBS: polarizing beam-splitter. PD: photodétector} \label{SetUp}
\end{figure}

The experiment uses metastable helium at room temperature. The $2^3\mathrm{S}_1 \rightarrow 2^3\mathrm{P}_1$ transition permits us to isolate a pure $\Lambda$ system involving only electronic spins and in which the Raman coherence lifetime is limited by the transit time of the atoms through the laser beam \cite{Gilles}. This transit time is lengthened thanks to non-dephasing collisions with ground-state atoms, leading to an effective Raman coherence lifetime of the order of 100~$\mu$s \cite{Goldfarb08}. Figure \ref{SetUp}(c) gives the schematic of the experimental set-up. The helium cell, filled with 1\,Torr of $^4$He, is 6\,cm long and has a diameter of 2.5\,cm. It is placed inside a three-layer $\mu$-metal shield for isolation from magnetic field inhomogeneities. The coupling and probe beams are derived from the same laser diode at 1.083 $\mu$m. The beam diameters are about 1 cm inside the cell. Helium atoms are excited to the metastable state by an RF discharge at 27 MHz. The coupling and probe beams are controlled in frequency and amplitude by two acousto-optic modulators (AOs), and recombined with a polarizing beam-splitter (PBS). The probe power is about 50\,$\mu$W and the coupling power can be varied between 0.5 and 22\,mW. A quarter-wave plate ($\lambda/4$) located at the entrance of the cell lets us alternate between orthogonal circular ($\sigma\perp\sigma$) and linear ($\mathrm{lin}\perp\mathrm{lin}$) polarizations. A variable longitudinal magnetic field (B) generated by a solenoid surrounding the helium cell lifts the degeneracy of the lower sublevels. The Landé factor is 2 for the ground state, leading to Zeeman shifts of $\pm$\,2.8\,kHz/mG for the $2^3\mathrm{S}_1,m= \pm 1$ levels. After the cell, polarization optics allows detection of only the probe.

In the usual configuration for EIT experiments along the $2^3\mathrm{S}_1 \rightarrow 2^3\mathrm{P}_1$ transition in $^4$He* \cite{Goldfarb08}, one uses circular polarizations for the pump and probe beams. Since the $m=0 \rightarrow m=0$ transition is forbidden, a $\sigma^+$ coupling beam pumps the atoms into the $m=+1$ ground state sublevel which is probed by a $\sigma^-$ beam [see Fig.\,\ref{SetUp}(a)]. In contrast, when the coupling beam is linearly polarized, atoms are equally pumped into both $m=\pm 1$ sublevels, which can then be probed by a perpendicular linearly polarized probe beam [see Fig.\,\ref{SetUp}(b)]. The experiments and simulations presented here compare these $\sigma\perp\sigma$ and $\mathrm{lin}\perp\mathrm{lin}$ configurations.

\begin{figure}[h]
  \includegraphics[width=1.0\linewidth]{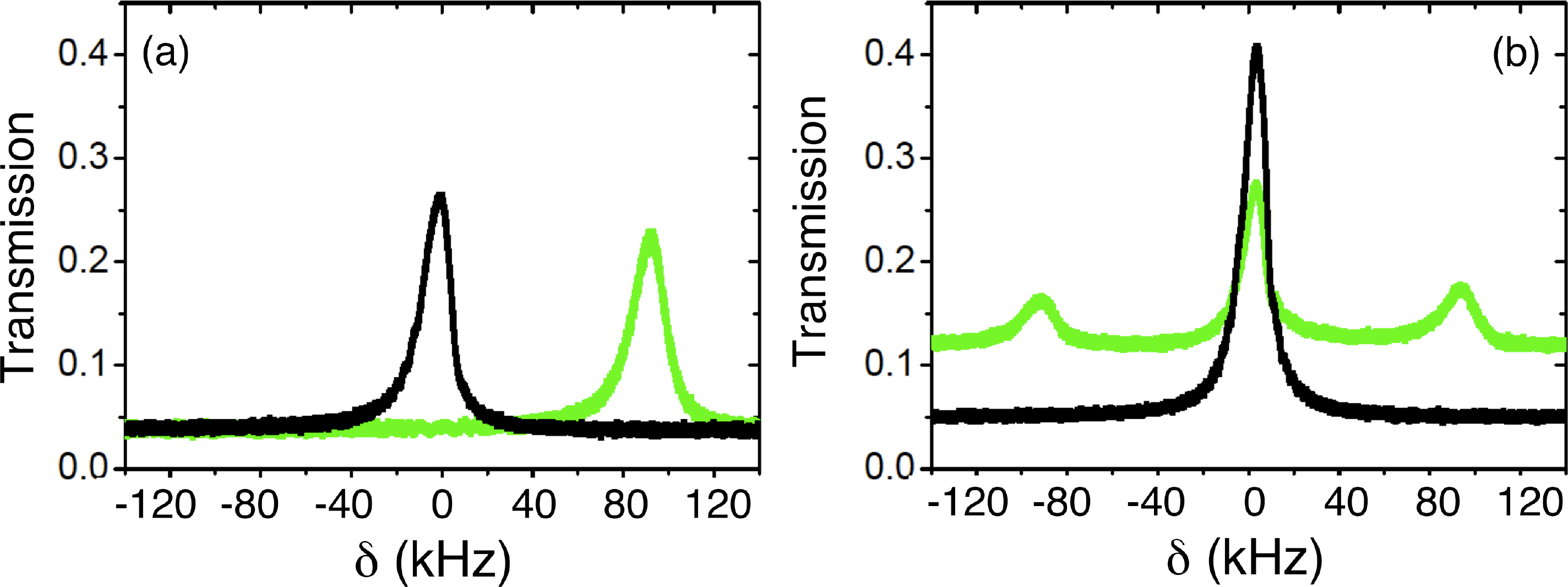}
  \caption{(color online) (a) Experimental results obtained with circular polarizations, with and without a 15\,mG magnetic field, which shifts the $m=\pm 1$ Zeeman sublevels by $\pm \Delta_\mathrm{Z}$. (b) Experimental results recorded with linear polarizations, with (green/grey) and without (black) the magnetic field. Two slightly off-resonance EIT peaks appear for $\delta=\pm 2\Delta_\mathrm{Z}$ ($\delta=\omega_{\mathrm{C}}-\omega_{\mathrm{P}}$), but the central resonance at $\delta=0$ does not correspond to any Raman resonance.} \label{ExperimentalResults}
\end{figure}

In the $\sigma\perp\sigma$ configuration, Fig.\,\ref{ExperimentalResults}(a) reproduces the experimentally recorded transmission spectrum of the probe versus $\delta$ with and without a magnetic field. As expected, the Zeeman shift $\Delta_\mathrm{Z}$ induced by the magnetic field shifts the EIT resonance. The Raman resonance is obtained for $\delta=\omega_{\mathrm{C}}-\omega_{\mathrm{P}}=2\Delta_\mathrm{Z}$, which is in perfect agreement with our data. Now, Fig.\,\ref{ExperimentalResults}(b) shows the results from the same experiment except that the pump and probe beams are now linearly and orthogonally polarized ($\mathrm{lin}\perp\mathrm{lin}$ configuration). Each beam thus excites equally the transitions $\Delta m=\pm 1$, as seen from Fig.\,\ref{SetUp}(b). Here, the two side peaks induced by the magnetic field occur at $\pm 2\Delta_Z$ and can be interpreted as EIT peaks. They appear because the pump beam couples both transitions with a slight optical detuning $\pm \Delta_Z$. In contrast, the central peak does not correspond to any Raman resonance. It occurs when the coupling and probe beams have the same frequency, while both transitions experience opposite frequency shifts induced by the applied magnetic field. The appearance of such a peak in the context of an EIT experiment is thus quite surprising.

The fact that this unexpected resonance occurs at zero frequency difference $\delta$ between the pump and the probe is reminiscent of CPO. However, the width of usual CPO resonances is given by the decay rate of the population of the upper level. In our case, since the \emph{lifetime of the excited level} is of the order of 100\,ns, this would lead to a resonance width of the order of 1\,MHz. It is clear from Fig.\,\ref{ExperimentalResults}(b) that the width of this extra central resonance is in the kHz range, and is quite close to the width of the side EIT resonances in Fig.\,\ref{ExperimentalResults}(b) or of those in Fig.\,\ref{ExperimentalResults}(a). Thus, if we want to interpret this extra resonance as due to population oscillations, it has to be related to the \emph{lifetime of the lower level}, which in our case is limited by the transit time and is thus compatible with the observed widths.

\begin{figure}[h]
  \includegraphics[width=1.0\linewidth]{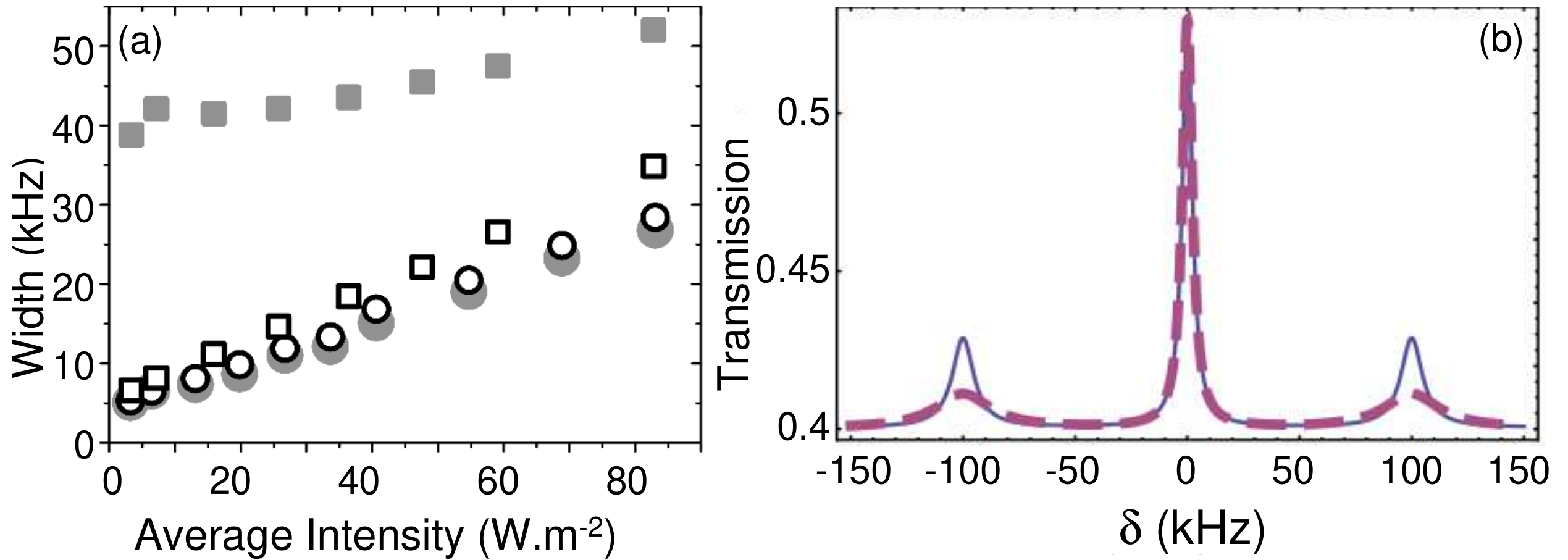}
  \caption{(color online) (a) Evolution of the resonance width versus average coupling intensity for the standard EIT resonance (squares) and for the central resonance in the $\mathrm{lin}\perp\mathrm{lin}$ configuration (circle). Open (filled) symbols correspond to measurements performed in the absence (presence) of a magnetic field gradient. (b) Simulated probe transmission spectra in the $\mathrm{lin}\perp\mathrm{lin}$ configuration for two values of the Raman coherence decay rate $\Gamma_\mathrm{R}/2\pi=3\,\mathrm{kHz}$ (continuous blue) and $\Gamma_\mathrm{R}/2\pi=12\,\mathrm{kHz}$ (dashed purple), obtained from a Floquet analysis of the three-level system (see text). The transit decay rate is kept fixed at $\Gamma_\mathrm{t}/2\pi=2\,\mathrm{kHz}$.}
  \label{BFieldEffect}
\end{figure}

A simple and effective way to check whether a resonance is due to population oscillations or not is to observe whether the introduction of a dephasing effect, which decreases the lifetime of the coherences, affects it or not. With this aim, we record the evolution of the widths of the different resonances with the coupling intensity, in the presence of an inhomogeneous magnetic field, simply realized by taking the cell partially out of the magnetic field shielding, and then again with proper magnetic shielding. The result is reproduced in Fig.\,\ref{BFieldEffect}(a). The standard EIT resonance is of course extremely sensitive to magnetic field gradients, because it relies on the Raman coherence. In contrast, the central resonance in the $\mathrm{lin}\perp\mathrm{lin}$ configuration here is totally unaffected by these gradients. This proves that this resonance is governed by the population lifetime in the lower state of the transition, and not by the Raman coherence lifetime.

We thus see a resonance due to population oscillations in the ground state. This resonance is observable here, and never in the case of CPO in a standard closed TLS, due to the fact that our three levels constitute two open TLSs. The extra resonance we get is thus linked to the extra resonances predicted by Friedmann {\it et al.} \cite{Friedmann} in 1986 in the case of four-wave mixing (FWM), when an intermediate decaying state is added to a TLS.

We could reproduce the experimental results by performing a first-order Floquet expansion of the density matrix of the three-level system in a manner similar to Wong {\it et al.} \cite{Wong}. Such simulated spectra are shown in Fig.\,\ref{BFieldEffect}(b) for two values of the Raman coherence linewidth. These confirm that the central resonance is not linked to the Raman coherences, unlike the EIT side peaks.

\begin{figure}[h]
  \includegraphics[width=1.0\linewidth]{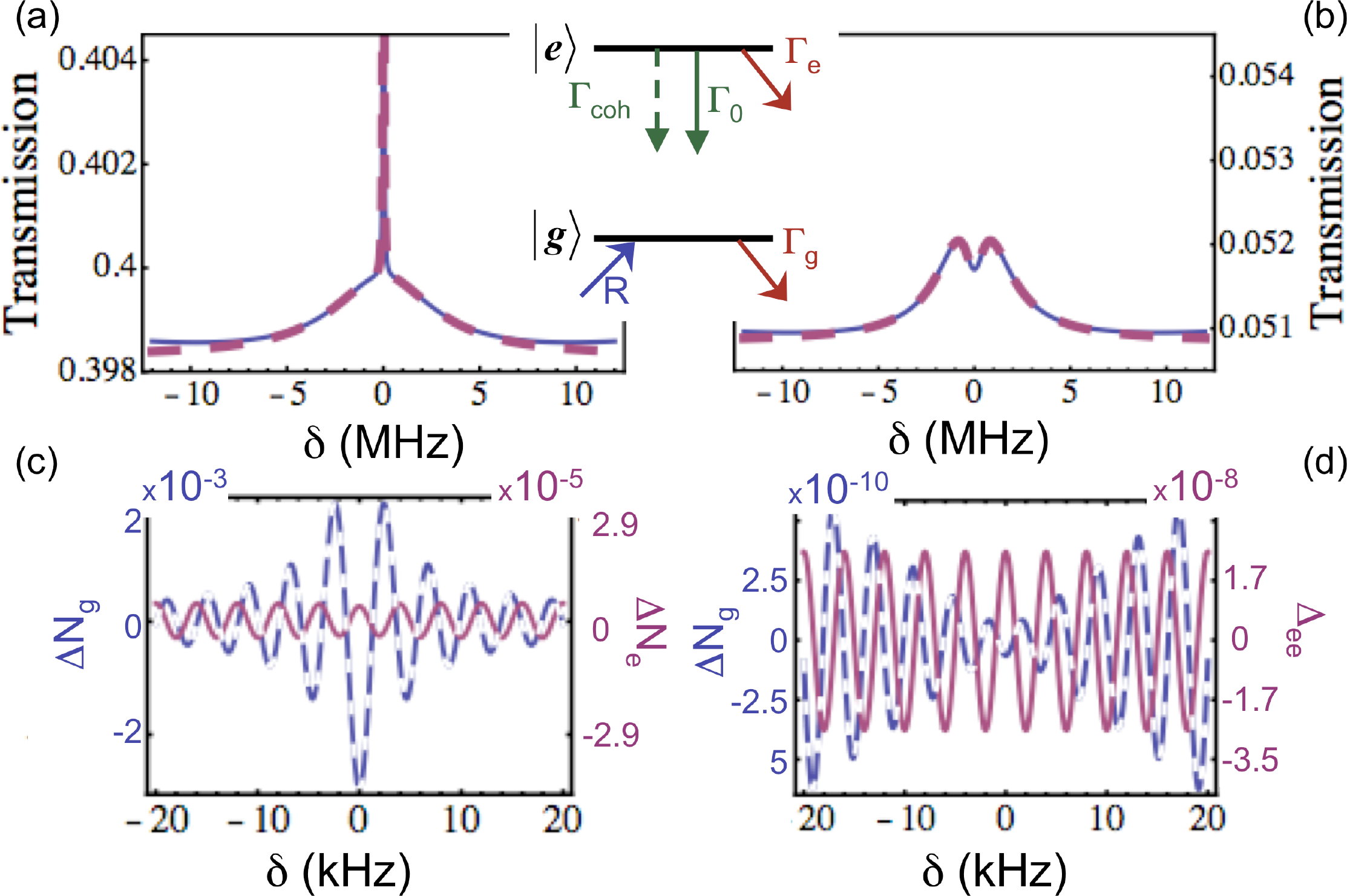}
  \caption{(color online) Simulations of the transmission of an open TLS in the case of (a) $\Gamma_{e}/2\pi=\Gamma_{0}/2\pi=1.6\,\mathrm{MHz}>\Gamma_g/2\pi=\Gamma_t /2\pi=2\,\mathrm{kHz}$, and (b) $\Gamma_e=\Gamma_t\ll \Gamma_{g}=\Gamma_{0}/2$. Continuous blue: simulations with the Floquet expansion of the complete density matrix. Dashed purple: simple rate equation model analysis. Inset: open TLS. $\Gamma_{\mathrm{coh}}$ is the optical coherence relaxation rate. (c), (d): Evolutions of the oscillating parts $\Delta N_e=(N_{1e}e^{-i\delta t}+\mathrm{c.c.})$ and $\Delta N_g=(N_{1g}e^{-i\delta t}+\mathrm{c.c.})$ of the excited (continuous purple) and ground (dashed blue) state populations versus $\delta$ at a fixed time $t=0.25\,\mathrm{ms}$. (c): same parameters as (a). (d): same parameters as (b).} \label{2levelSimulations}
\end{figure}
If the central resonance is, as we suspect, due to population oscillations in the lower state sublevels, we should be able to reproduce the experimental results in the rate equation approximation, i.e., with all the coherences adiabatically eliminated. Using the notations defined in the caption of Fig.\,\ref{2levelSimulations}, we model one of the legs of the $\Lambda$ system (either the $\sigma^+$ or the $\sigma^-$ transition) as an open TLS, the other leg of the $\Lambda$ playing the role of an extra decay channel for the upper level. The rate equations for the populations $N_e$ and $N_g$ in the upper and lower levels $\left|e\right\rangle$ and $\left|g\right\rangle$ are then given by
\begin{eqnarray}
	\frac{\mathrm{d}N_e}{\mathrm{d}t}&=&-(\Gamma_0+\Gamma_e)N_e + \frac{I}{\hbar\omega}\sigma(N_g-N_e) ,\\
	\frac{\mathrm{d}N_g}{\mathrm{d}t}&=&\Gamma_0 N_e+ R -\Gamma_g N_g - \frac{I}{\hbar\omega}\sigma(N_g-N_e) ,
\end{eqnarray}
where $\sigma$ is the absorption cross section, and $\Gamma_0$, $\Gamma_e$, and $\Gamma_g$ are the population decay rates of the excited state to the ground one, the excited state to other states, and the ground state to other states, respectively. In the presence of a pump and a probe beam with a frequency difference $\delta$, the total intensity reads $I=I_0+(I_1e^{-i\delta t}+\mathrm{c.c.})$ with $I_1\ll I_0$. $R$ is the feeding rate of the lower level. At first order, the populations are expanded as $N_j(t)=N_{0j}+(N_{1j}e^{-i\delta t}+\mathrm{c.c.})$, where $j=g,e$. Then the oscillating part of the population inversion $W_1=N_{1e}-N_{1g}$ is found to be given by
\begin{equation}
W_1=\frac{-W_0I_1\Gamma_0(\Gamma_e+\Gamma_g-2i\delta)}{I_{\mathrm{sat}}(\Gamma_0+\Gamma_e-i\delta)(\Gamma_g-i\delta)+I_0\Gamma_0(\Gamma_e+\Gamma_g-2i\delta)}\label{Eq03} ,
\end{equation}
where
\begin{equation}
W_0=-\frac{R I_{\mathrm{sat}}(\Gamma_0+\Gamma_e)}{I_{\mathrm{sat}}\Gamma_g(\Gamma_0+\Gamma_e)+I_0\Gamma_0(\Gamma_g+\Gamma_e)}
\end{equation}
is the DC part of the population inversion, and $I_{\mathrm{sat}}=\Gamma_0\hbar\omega/\sigma$ is the saturation intensity. When $\Gamma_e=\Gamma_g=0$, Eq.\,(\ref{Eq03}) reduces to the usual CPO resonance with a width given by $\Gamma_0$ \cite{Piredda}. In contrast, when $\Gamma_g \ll \Gamma_e$, assuming also that $\delta \ll \Gamma_0,\Gamma_e$, Eq.\,(\ref{Eq03}) reduces to:
\begin{equation}
W_1=-W_0 \frac{I_1}{I_{\mathrm{sat}}}\frac{\Gamma_0\Gamma_e}{\Gamma_0+ \Gamma_e}\left(\frac{1}{\Gamma_g + \frac{\Gamma_0\Gamma_e}{\Gamma_0+\Gamma_e}\frac{I_0}{I_{\mathrm{sat}}}-i\delta}\right) \label{Eq05}\  .
\end{equation}
Equation (\ref{Eq05}) thus exhibits a resonance at $\delta=0$, with a width limited by $\Gamma_g$ at vanishing coupling intensities $I_0\ll I_{\mathrm{sat}}$. Figure \ref{2levelSimulations}(a) reproduces the corresponding simulation results, in which $R$ is taken equal to the transit rate $\Gamma_t$, $\Gamma_g=\Gamma_t$, and $\Gamma_{e}=\Gamma_0 \gg \Gamma_g$: each TLS along each leg of the $\Lambda$ system gives a resonance with a kHz width limited by the transit time of the atoms through the beam. Our simple rate equation model (dashed purple line in Fig.\,\ref{2levelSimulations}(a)) is in very good agreement with a simulation based on a first-order Floquet expansion of the full density matrix (continuous blue line in Fig.\,\ref{2levelSimulations}(a)), showing the validity of the explanation of the extra resonance in terms of lower level CPO. This is confirmed by Fig.\,\ref{2levelSimulations}(c), which shows the evolutions of the oscillating parts $\Delta N_e=2\;\mathrm{Re}(N_{1e}e^{-i\delta t})$ and $\Delta N_g=2\;\mathrm{Re}(N_{1g}e^{-i\delta t})$ of the excited and ground state populations at a fixed time $t$. The oscillations of the lower level population exhibit a resonance with a few kHz width, while the oscillations of the upper level population have a much smaller amplitude and exhibit no visible resonance at this frequency scale.

It is worth noticing that, in the opposite case where $\Gamma_g \gg \Gamma_e$, one predicts the existence of a transmission dip of width $\Gamma_e$ (see Fig.\,\ref{2levelSimulations}(b)). Similar subnatural absorption features were discussed both theoretically and experimentally by different groups at the end of the eighties and the beginning of the nineties \cite{Berman, Wilson-Gordon}, with an emphasis on FWM, considerably different from the point of view developed here. Figure \ref{2levelSimulations}(d) proves that the narrow dip of Fig.\,\ref{2levelSimulations}(b) is due to a decrease of the amplitude of the oscillations of the lower level population.

\begin{figure}[h]
  \includegraphics[width=1.0\linewidth]{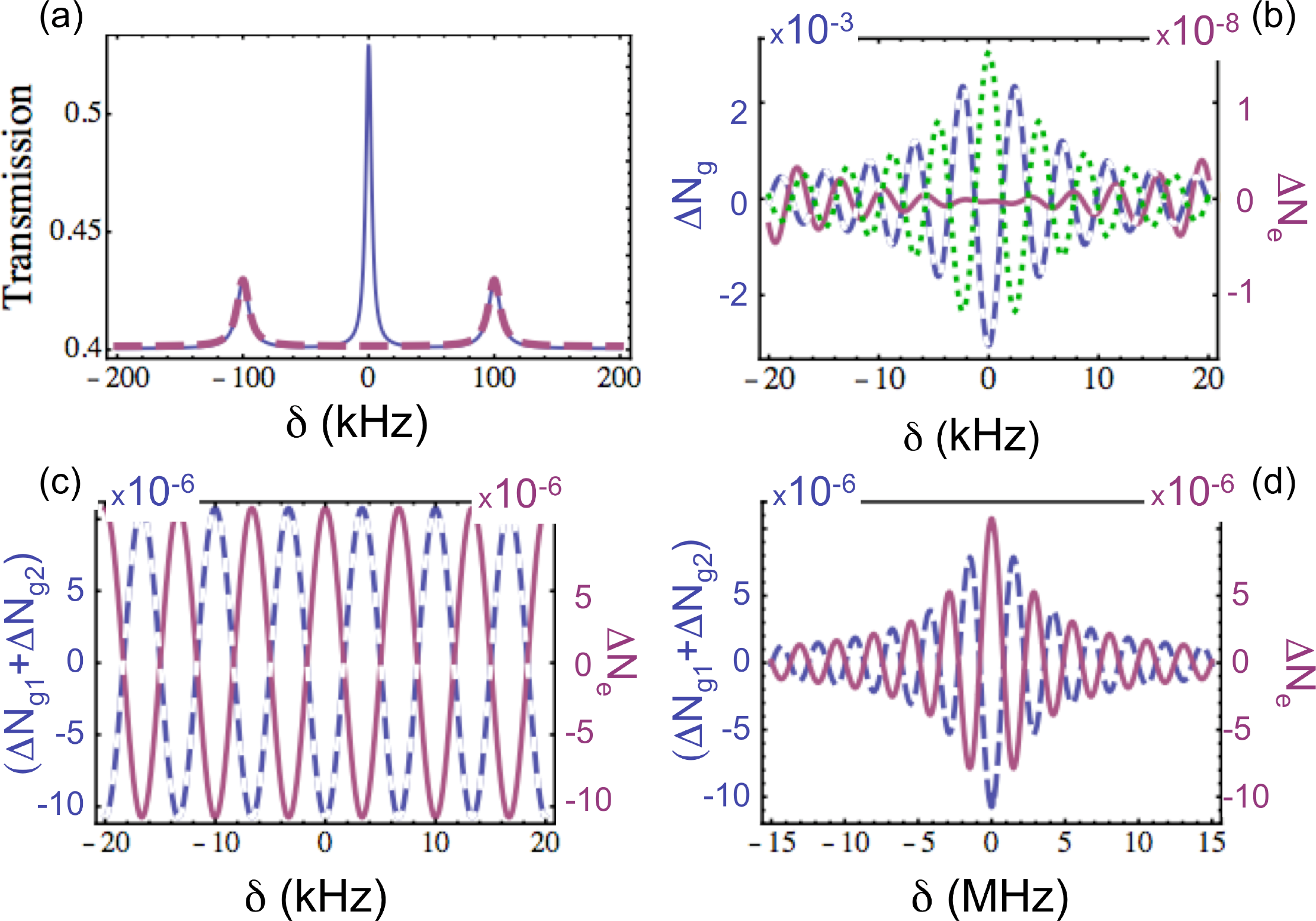}
  \caption{(color online) (a) Simulated probe transmission spectrum in a $\Lambda$ system in the $\mathrm{lin}\perp\mathrm{lin}$ (continuous blue) and $\mathrm{lin}\parallel\mathrm{lin}$ (dashed purple) configurations. (b) Evolutions versus $\delta$ of the oscillating parts $\Delta N_e=(N_{1e}e^{-i\delta t}+\mathrm{c.c.})$ of the excited state (continuous purple), and $\Delta N_{g1}=(N_{1g1}e^{-i\delta t}+\mathrm{c.c.})$ (dashed blue) and $\Delta N_{g2}=(N_{1g2}e^{-i\delta t}+\mathrm{c.c.})$ (dotted green) of the two ground states, at a fixed time $t$ in the $\mathrm{lin}\perp\mathrm{lin}$ configuration. (c),(d) Evolutions versus $\delta$ of the oscillating part $\Delta N_e$ of the excited state (continuous purple) and of the sum $\Delta N_{g1}+\Delta N_{g2}$ of the oscillating parts of the two ground state populations (dashed blue), in the $\mathrm{lin}\parallel\mathrm{lin}$ configuration.} \label{3levelSimulations}
\end{figure}

However, our $\Lambda$ system is more than an open TLS: it consists of two open TLSs which are interdependent and excited by the same pair of coupling and probe fields. One could thus expect the lower level CPO of these two TLSs to interfere constructively or destructively, depending on the relative signs of their excitation fields. This is why the results of Fig.\,\ref{3levelSimulations} permit us to compare the results of the Floquet simulations in our $\Lambda$ system for perpendicular ($\mathrm{lin}\perp\mathrm{lin}$) or parallel ($\mathrm{lin}\parallel\mathrm{lin}$) linearly polarized pump and probe beams.
%The only difference between the two cases lies in the fact that there is a minus sign in the expansion of one of the linear polarizations in terms of circular polarizations [namely, $\mathbf{\hat{y}}=(\mathbf{\hat{\sigma^+}}-\mathbf{\hat{\sigma^+}})/2i$], which is absent in the other one [$\mathbf{\hat{x}}=(\mathbf{\hat{\sigma^+}}+\mathbf{\hat{\sigma^+}})/2$].
Figure \ref{3levelSimulations}(a) shows that the ground state CPO resonance is present only in the $\mathrm{lin}\perp\mathrm{lin}$ configuration, but disappears in the $\mathrm{lin}\parallel\mathrm{lin}$ configuration.

Indeed, in the $\mathrm{lin}\perp\mathrm{lin}$ configuration (see Fig.\,\ref{3levelSimulations}(b)), the populations of the two ground state sublevels oscillate in antiphase, due to the fact that they are driven by intensity modulations which are in antiphase. The two CPO resonances induced by the two legs of the $\Lambda$ then add constructively, giving birth to a sharp resonance limited by the decay rate of the populations of the ground states. In contrast, in the $\mathrm{lin}\parallel\mathrm{lin}$ configuration, the populations of the two ground states oscillate in phase (see Fig.\,\ref{3levelSimulations}(c)), and the system behaves like a closed TLS, showing no resonance in the kHz range. Of course, as shown in Fig.\,\ref{3levelSimulations}(d), the system can still exhibit the much broader, usual CPO resonance with a width given by $\Gamma_0$, just like an ordinary closed TLS.

It is noted that similar resonances have appeared in previous works, both experimental and theoretical \cite{Lezama, Lipsich}. However, these papers focused on electromagnetically induced absorption and did not discuss the existence of such sharp CPO-induced transmission windows.

The interest in CPO has been lasting for more than a decade since many people work on possible applications of slow and fast light, in particular, for microwave photonics \cite{Ohman07,Xue09_2,Shumakher09_2,Berger11}. Besides, recent theoretical proposals suggest use of such long-lived CPOs even for applications in spatial optical memories \cite{Eilam} or narrowband biphoton generation \cite{Sharypov}. The system described here is of interest as it can be made to behave as either a closed or an open TLS by a simple change of the polarization direction of one of the coupling and probe beams. It also has the advantage of exhibiting conservation of its total population, avoiding the need for a repumping laser. Moreover, the decay rate of the ground state population is very long and is not limited by spontaneous emission, but only by the transit of the atoms through the laser beam. Other $\Lambda$ systems in solids (such as rare-earth ions or NV centers in diamonds) might even go beyond the present limitation set by the transit time. Such coupled open TLSs are thus good candidates for the experimental implementation of the recent theoretical proposals based on narrow CPO effects.

The authors thank Amrita Madan for interesting discussions, and acknowledge partial support from the Agence Nationale de la Recherche, the Triangle de la Physique, and the Indo-French Center for the Promotion of Advanced Research (IFCPAR/CEFIPRA).

\end{document}